\documentclass[aps,onecolumn,pra,superscriptaddress]{revtex4}

\usepackage{graphicx}
\usepackage{subfigure}
\usepackage{comment}
\usepackage{dcolumn}
\usepackage{bm}
\usepackage{amsmath,amssymb}
\usepackage{amsthm}
\usepackage{mathrsfs}
\usepackage{indentfirst}
\usepackage{float}
\usepackage{braket}
\usepackage[utf8]{inputenc}

\usepackage{tikz}
\usetikzlibrary{arrows, shapes.gates.logic.US, calc}
\usetikzlibrary{angles,quotes}
\tikzstyle{branch}=[fill, shape=circle, minimum size=3pt, inner sep=0pt]
\usepackage{blochsphere}
\usepackage{mathtools}
\usepackage{color,graphicx} 
\newcommand\D{\operatorname{d}\!}
\newcommand{\rh}[1]{\mathrm{\hat{#1}}}

\newcommand{\cl}[1]{\mathcal{#1}}
\newcommand{\xv}{\textbf{x}}
\newcommand{\yv}{\textbf{y}}
\newcommand{\kv}{\textbf{k}}
\newcommand{\pvec}{\textbf{p}}
\newcommand{\J}{{\rh{\textbf{J}}}}
\newcommand{\update}[1]{}
\newcommand{\rC}{r_\text{\tiny C}}
\newcommand{\RC}{R_\text{\tiny C}}
\newcommand{\kB}{k_\text{\tiny B}}
\newcommand{\Tr}{\operatorname{Tr}}

\usepackage{bbold}

\usetikzlibrary{quantikz}

\usepackage{hyperref}
\hypersetup{
 colorlinks=true,
 citecolor=red,
 linkcolor=blue,
 urlcolor=blue,
 pdfpagemode=UseNone,
 pdfstartview=FitH}

\newcommand{\orcidicon}[1]{\href{https://orcid.org/#1}{\includegraphics[height=\fontcharht\font`\B]{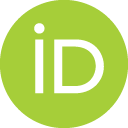}}}

\usepackage{subfiles}

\begin{document}

\title{Experimental bounds on linear-friction dissipative collapse models from levitated optomechanics}

\author{Giovanni Di Bartolomeo\,\orcidicon{0000-0002-1792-7043}}
\email{giovanni.dibartolomeo@phd.units.it}
\affiliation{Department of Physics, University of Trieste, Strada Costiera 11, 34151 Trieste, Italy}
\affiliation{Istituto Nazionale di Fisica Nucleare, Trieste Section, Via Valerio 2, 34127 Trieste, Italy}
\author{Matteo Carlesso\,\orcidicon{0000-0002-9929-7291}}
\affiliation{Department of Physics, University of Trieste, Strada Costiera 11, 34151 Trieste, Italy}
\affiliation{Istituto Nazionale di Fisica Nucleare, Trieste Section, Via Valerio 2, 34127 Trieste, Italy}
\affiliation{Centre for Quantum Materials and Technologies,
School of Mathematics and Physics, Queens University, Belfast BT7 1NN, United Kingdom}

\begin{abstract}
Collapse models constitute an alternative to quantum mechanics that solve the well-know quantum measurement problem. In this framework,
a novel approach to include dissipation in collapse models has been recently proposed, and  awaits experimental scrutiny. Our work establishes experimental bounds on the so-constructed linear-friction dissipative Diósi-Penrose (dDP) and Continuous Spontaneous localisation (dCSL) models by exploiting experiments in the field of levitated optomechanics. Our results in the dDP case exclude collapse temperatures below $ 10^{-13}$\,K and $ 6 \times 10^{-12}$\,K respectively for values of the localisation length smaller than $10^{-6}$\,m and  $10^{-8}$\,m. In the dCSL case the entire parameter space is excluded for values of the temperature lower than $ 6 \times 10^{-9}$\,K.
\end{abstract}

\maketitle

\section{Introduction}\label{intro}
Models of spontaneous wavefunction collapse, or simply collapse models \cite{bassi2003dynamical,bassi2013models,carlesso2022present}, represent a well established paradigm  in the realm of quantum foundations, and constitute a strong figure of merit for the study of the macroscopic limits of quantum mechanics. Indeed, their investigation finds motivation in the lack of observed quantum superpositions at the macroscopic scale: while quantum mechanics has proven highly successful in describing microscopic phenomena, it has yet to explain why macroscopic objects do not exhibit quantum superpositions although the theory predicts them. The key idea of collapse models is that quantum mechanics must be modified to explain  the quantum-to-classical transition at macroscopic scales. Thus, they add suitably constructed phenomenological terms to the standard Schr\"odinger equation. Their action can be seen as that of a noise field that leads to the collapse of the wavefunction. Depending on the specific collapse model, the origin of such a field can be either of unknown origin or be related to the gravitational field. These models introduce additional free parameters that control the collapse mechanism, and their validity is subject to experimental verification. 
Although interferometric experiments, where a superposition is directly probed, face increasing challenges that grow with the system size, non-interferometric ones play a crucial role in testing collapse models \cite{Bilardello:2016aa,vinante2016upper,
PhysRevD.94.124036,
PhysRevD.99.103001,
vinante2017improved,helou2017lisa,
PhysRevLett.125.100404,
PhysRevResearch.2.013057,PhysRevResearch.2.023349,
donadi2021novel,donadi2021underground,
PhysRevLett.129.080401}.  Such experiments focus on monitoring quantities like position or energy, and are relatively easier to perform. They are able to provide strong bounds in the parameter space of specific collapse models. For an overview on the state-of-the-art in theory and experiments, the reader can refer to Ref.~\cite{carlesso2022present}.

The two most studied collapse models are the Diósi-Penrose (DP) model \cite{diosi1987universal,penrose1996gravity}  and the Continuous Spontaneous Localisation (CSL) model \cite{pearle1989combining,ghirardi1990markov}. The latter is parametrised by a collapse rate $\lambda$ and a localisation length $\rC$. Conversely, the former model, which is related to gravity, is characterised only by a localisation length $R_0$ as the collapse rate is fixed by the gravitation constant $G$. An acknowledged challenge within collapse models is the energy divergence due to the collapse mechanism. Although the rate of energy increase is extremely small, e.g. for the CSL this is of the order of $10^{-15}$\,K/yr for a free nucleon with $\lambda=10^{-16}$\,s$^{-1}$ at $\rC=10^{-7}\,$m \cite{bassi2003dynamical}, and the interaction with the external noise field is expected to violate energy conservation solely for the system, one does not expect that the noise field will convey energy indefinitely to the system. To address this concern, dissipative extensions of collapse models were proposed as a solution \cite{smirne2015dissipative,bahrami2014role}, implying the existence of a fundamental and universal damping mechanism which can be probed by mechanical systems with very low dissipation \cite{nobakht2018unitary,PhysRevResearch.2.023349,vinante2019testing}.

Recently, a new approach to the introduction of dissipation in collapse model has been proposed \cite{PhysRevA.108.012202}. The latter is based on a different mechanism with respect to that previously proposed, namely the linear-friction of the current of the many-body system. Such an approach has not been tested yet. The present work falls within this context.

We derive the first experimental bounds on linear-friction dissipative DP (dDP) and CSL (dCSL) models from levitated optomechanics characterised by ultralow damping \cite{PhysRevResearch.2.023349,vinante2019testing,dania2023ultra}. In particular, for the dDP model, values of the temperature $T_\beta$ of the collapse field lower than $ 10^{-13}$\,K and $ 6 \times 10^{-12}$\,K are excluded respectively for values of the localisation length smaller than $10^{-6}$\,m and  $10^{-8}$\,m. On the other hand, for the dCSL model,  the entire parameter space is excluded for $T_\beta$ lower than $ 6 \times 10^{-9}$\,K. Finally, we compare the approach recently proposed in \cite{PhysRevA.108.012202} with those previously suggested in \cite{smirne2015dissipative,bahrami2014role}, and conclude that they can be in principle experimentally distinguished.

\section{The model}\label{model}
Here, we briefly present the universal dissipative mechanism for collapse models of many-body systems, which was introduced in Ref.~\cite{PhysRevA.108.012202}.  We start from the following master equation
\begin{equation}
\label{ME}
\frac{\D}{\D t}\hat{\rho}_t=-\frac{i}{\hbar}[\hat{H},\hat{\rho}_t]+\cl{D}\hat{\rho}_{t},
\end{equation}
where $\hat{H}$ is the Hamiltonian of the system and 
\begin{equation}
\label{dissipator}
\begin{aligned}
\cl{D}\hat{\rho}_t &= \frac{1}{\hbar^2}\int\D^{3}x\int \D^{3}y\, D(\xv-\yv) \left(\hat{L}(\xv)\hat{\rho}_t\hat{L}^{\dagger}(\yv)-\frac12\{\hat{L}^{\dagger}(\xv)\hat{L}(\yv),\hat{\rho}_t\}\right).
\end{aligned}
\end{equation}
The dissipation is introduced in the model by considering the following Lindblad operator $\hat{L}(\xv)$ 
\begin{equation}
\label{ansatz}
\hat{L}(\xv)=\rh{\mu}(\xv)-i\frac{\hbar\beta}{4}\nabla_\xv \J(\xv),
\end{equation}
where $\beta$ is a free parameter driving the dissipation mechanism. A similar method has been
considered for a  gravity-related model in \cite{PhysRevD.104.104027}. When $\beta$ is set to zero, one obtains the standard (non-dissipative) collapse master equation. The mass density $\rh{\mu}(\xv)$ and the current $\J(\xv)$ in the second-quantization framework respectively read as
\begin{subequations}
\begin{align}
\label{mass_current_second_q}
& \rh{\mu}(\xv) = m \hat{\psi}^{\dagger}(\xv)\hat{\psi}(\xv),\\
&\J(\xv)=-i\frac{\hbar}{2}\left(\hat{\psi}^{\dagger}(\xv)\nabla_\xv\hat{\psi}(\xv)-\nabla_\xv\hat{\psi}^{\dagger}(\xv)\hat{\psi}(\xv)\right),
\end{align}
\end{subequations}
with $\hat{\psi}(\xv)$ being the (fermionic) annihilation field operator. In the following sections we will work in the first-quantization. Thus, for a system of $N$ point-like particles of mass $m$, the mass density and the current can be expressed as
\begin{equation}
%\begin{align}
\label{mass_current_firstq}
 \rh{\mu}(\xv) = m \sum_{j=1}^{N}\delta(\xv-\hat{\xv}_{j}),\quad
\J(\xv)= \frac{1}{2}\sum_{j=1}^{N}\Bigl\{\hat{\pvec}_{j}, \delta(\xv-\hat{\xv}_{j})\Bigr\},
%\end{align}
\end{equation}
where $\hat{\xv}_{j}$ and $\hat{\pvec}_{j}$ are respectively the position and momentum operator of the j-th particle.

The form of the kernel $D(\xv-\yv)=(2\pi)^{-3}\int{\D^3k} \,D_\kv e^{i \kv (\xv-\yv)}$ in Eq.~\eqref{dissipator} depends on the specific collapse model. Here we consider the dissipative Diósi-Penrose (dDP) \cite{diosi1987universal,penrose1996gravity} and the dissipative Continuous Spontaneous localisation (dCSL) \cite{pearle1989combining,ghirardi1990markov} models, which correspond respectively to
\begin{equation}
D_\kv =\exp(-\sigma^2 k^2)\times
\begin{cases}
\hbar^2\gamma & (\text{CSL}),\\
4\pi\hbar G/k^2 & (\text{DP}),
\end{cases}
\end{equation}
 where $k=|\kv|$. Here,  the $k^{-2}$ term in the second expression comes from the Fourier transform of the Newtonian potential  $V(\xv-\yv) = -G/|\xv-\yv|$. In the DP model the decoherence rate is set by the Newton constant $G$ and $\sigma = R_{0}$ is a free parameter representing the spatial cut-off due to the regularization procedure. The CSL model can be described in terms of two free parameters being $\gamma = (\sqrt{4\pi}\sigma)^3 \lambda / m_0^2$ and $\sigma = \rC$, which are respectively the collapse rate and localisation length of the model ($m_{0}$ is a reference mass chosen as that of a nucleon).

In the upcoming sections, we will delve into the dissipative dynamics of the center of mass of a one-dimensional mechanical oscillator. Specifically, we will analyze the dynamics of the center of mass of a rigid body composed of $N$ particles. After having suitably linearised the dynamics,  we derive the modified Langevin equations describing the dynamics of the mechanical oscillator.

\section{Dissipative dynamics of the center of mass of a $N$-particle system}\label{center_of_mass}
We compute the  dynamics of the center of mass of a rigid system made of $N$ particles. For convenience, we rewrite Eq.~\eqref{dissipator} in the Fourier representation
\begin{equation}
\label{dissipator_fourier}
%\begin{aligned}
\cl{D}\hat{\rho}_t = \frac{1}{\hbar^2}\int\frac{\D^{3}k}{(2\pi)^3} D_\kv 
\left(\hat{L}_\kv\hat{\rho}_t\hat{L}_{\kv}^{\dagger}-\frac12\{\hat{L}_{\kv}^{\dagger}\hat{L}_\kv,\hat{\rho}_t\}\right),
%\end{aligned}
\end{equation}
where now the Lindblad operators become
\begin{equation}
\label{ansatz_fourier}
\hat{L}_\kv=\rh{\mu}_\kv +\frac{\hbar\beta}{4}\kv \J_\kv,
\end{equation}
and the Fourier representation of the mass density and of the current is
\begin{equation}
\label{mass_current_fourier}
\rh{\mu}_\kv = m \sum_{j=1}^{N}e^{i\kv\hat{\xv}_{j}},\quad \J_\kv= \frac{1}{2}\sum_{j=1}^{N}\left\{\hat{\pvec}_{j}, e^{i\kv\hat{\xv}_{j}}\right\}.
\end{equation}
In general the position and momentum operators in Eq.~\eqref{mass_current_fourier} can be written as 
\begin{equation}
\label{coordinates}
\hat{\xv}_j = \hat{\xv} +\xv_j^{(0)} +  \Delta\hat{\xv}_{j},\quad \hat{\pvec}_j = \frac{m}{M}\hat{\pvec} +\pvec_j^{(0)} +  \Delta\hat{\pvec}_{j},
\end{equation}
where  $\hat{\xv}$ and $\hat{\pvec}$ are the position and momentum operators of the center of mass, $\xv_j^{(0)}$ and $\pvec_j^{(0)}$ are the classical equilibrium position and momentum of the $j$-th particle with respect to the center of mass, and $\Delta\hat{\xv}_{j}$ and $\Delta\hat{\pvec}_{j}$ are the relative fluctuations, and $M$ is the total mass of the system. Under the assumption of a rigid body, the relative fluctuations are negligible, namely $\Delta\hat{\xv}_{j} = \Delta\hat{\pvec}_{j}= 0$. By substituting Eq.~\eqref{coordinates} into Eq.~\eqref{mass_current_fourier} and by assuming that the spread of the wavefunction of the center of mass is much smaller than $\sigma$, we can Taylor expand the mass density and the current for small fluctuations of $\hat{\xv}$, finding
\begin{subequations}
\begin{align}
&\hat{\mu}_\kv\simeq \mu_\kv + i \kv \mu_\kv\hat{\xv},\\
& \J_\kv \simeq \textbf{J}_\kv + \frac{\mu_\kv}{M}\hat{\pvec}+ i \kv \textbf{J}_\kv \hat{\xv} + \frac{i\kv}{2M}\mu_\kv\{\hat{\pvec},\hat{\xv}\},
\end{align}
\end{subequations}
where $\mu_\kv = m \sum_{j=0}^{N}e^{i\kv\xv^{(0)}_j}$ and $\textbf{J}_\kv = \sum_{j=0}^{N}\pvec_{j}^{(0)}e^{i\kv\xv_{j}^{(0)}}$ are respectively the classical mass density and current of the system in the Fourier representation. We notice that for a rigid body $\pvec_j^{(0)}=0$ thus $\textbf{J}_\kv = 0$.

For the sake of simplicity, we reduce the problem in one dimension, namely $\hat{\xv} = (\hat{x},0,0)$ and $\hat{\pvec} = (\hat{p},0,0)$. Moreover, by assuming small $\kv$ (i.e., $|\kv| \ll 1/\sigma$) we neglect all the terms of order higher than $O(\kv^{2})$ and substitute the latter expressions for the mass density and the current in Eq.~\eqref{dissipator_fourier}. In such a way, we obtain the following master equation for the motion of the center of mass in the linear limit
\begin{equation}
\label{final_linear_ME}
\begin{aligned}
\frac{\D}{\D t}\hat{\rho}_{cm}=-\frac{i}{\hbar}[\hat{H},\hat{\rho}_{cm}]+\eta\left(\hat{L}\hat{\rho}_{cm}\hat{L}^{\dagger}-\frac{1}{2}\{\hat{L}^{\dagger}\hat{L},\hat{\rho}_{cm}\}\right),
\end{aligned}
\end{equation}
where $\hat{L} = \hat{x}+i\alpha\hat{p}$ with $\alpha = {\Gamma}/{2\eta\hbar}$ 
and where

\begin{equation}
\label{gamma_eta}
\Gamma = \frac{\hbar^{2}\beta\eta}{2M},\quad
\eta = \frac{1}{\hbar^{2}}\int \frac{\D^{3}k}{(2\pi)^3}k_{x}^{2}D_{\kv}|\mu_{\kv}|^{2},
\end{equation}
are respectively the dissipation and the diffusion rates with $\kv =(k_x,k_y,k_z)$. For the purpose of this work, we can consider the case of a system being a continuous and homogeneous sphere of radius $r$. For such a case, we have
\begin{equation}
\mu(\xv) = \frac{3M}{4\pi r^3}\theta(r-|\xv|),
\end{equation}
where $\theta$ is the Heaviside function. Then, $\eta$ takes the following form 
\begin{equation}
\label{etaDP}
\begin{aligned}
\eta_\text{DP} = &\frac{G M^2R_0}{\hbar \sqrt{\pi} r^6}\left[-3 r^2 + 2 R_0^2 + e^{-(r^2/R_0^2)} (r^2 - 2 R_0^2)+\sqrt{\pi} r^3 \operatorname{erf}(r/R_0)\right],
\end{aligned}
\end{equation}
for the dDP model and 
\begin{equation}
\label{etaCSL}
\begin{aligned}
\eta_\text{CSL} = \lambda\frac{3 e^{-(r^2/
  \rC^2)} M^2 \rC^2}{m_0^2 r^6}\left[r^2 + 2 \rC^2 + 
   e^{r^2/\rC^2} (r^2 - 2 \rC^2)\right],
\end{aligned}
\end{equation}
for the dCSL model.

\section{Application to Langevin equations of a mechanical oscillator}\label{oscillator}
In the present section, we explore the dissipative dynamics of the center of mass of a one-dimensional mechanical oscillator in order to set experimental bounds on dDP and dCSL free parameters. To include the effects of dissipation in the Langevin equations for the mechanical oscillator \cite{mancini1994quantum}, we use the following unitary stochastic unravelling \cite{gardiner2004quantum,nobakht2018unitary} of Eq.~\eqref{final_linear_ME}
\begin{equation}
\label{SDE}
\D \ket{\psi_t} = \left[-\frac{i}{\hbar}\hat{H}\D t + \hat{L}\D \hat{B}^{\dagger}_t- \hat{L}^{\dagger}\D \hat{B}_t-\frac{\eta}{2}\hat{L^{\dagger}}\hat{L} \D t\right]\ket{\psi_t},
\end{equation}
where $\hat{B}_t$ is called quantum noise and is equipped with the following statistical features: $\mathbb E[\D \hat{B}_t]=0$, $\mathbb{E}[\D \hat{B}_t\D \hat{B}^{\dagger}_t] = \eta\D t$ and $\mathbb{E}[\D \hat{B}^{\dagger}_t\D \hat{B}^{\dagger}_t] = \mathbb{E}[\D \hat{B}_t\D \hat{B}_t] = \mathbb{E}[\D \hat{B}^{\dagger}_t\D \hat{B}_t] = 0$. 
From Eq.~\eqref{SDE}, one can build the Langevin equation for a generic operator $\hat{O}$ via
\begin{equation}
\label{langevin}
\begin{aligned}
\frac{\D}{\D t}\hat{O} &= \frac{i}{\hbar}[\hat{H},\hat{O}]+\eta\left(\hat{L}^{\dagger}\hat{O}\hat{L}-\frac{1}{2}\{\hat{L}^{\dagger}\hat{L},\hat{O}\}\right)+\hat{b}_t^{\dagger}[\hat{O},\hat{L}]+\hat{b}_t[\hat{L}^{\dagger},\hat{O}],
\end{aligned}
\end{equation}
where $\hat{b}_t = \D \rh{B}_t/\D t$.

Finally, by using Eq.~\eqref{langevin} with $\hat{O} = \hat{x}, \hat{p}$ and $\hat{H}=\hat{p}^2/2M+M\omega_0\hat{x}^2/2$, we find the following modified Langevin equations for a one-dimensional mechanical oscillator of mass $M$ and frequency $\omega_0$ 
\begin{subequations}
\begin{align}
\frac{\D \hat{x}}{\D t}&= \frac{\hat{p}}{M}-\frac{\Gamma}{2}\hat{x}-\hbar \alpha\hat{w}_x,\label{opto_langevin_a}\\
\frac{\D \hat{p}}{\D t}&= -M\omega_0^2\hat{x}
-\Bigl(\frac{\Gamma}{2} +\gamma_m\Bigr)\hat{p}+\xi-\hbar \hat{w}_p,\label{opto_langevin_b}
\end{align}
\end{subequations}
where $\hat{x}$ is the position operator for the center of mass of the oscillator, $\hat{p}$ is the corresponding momentum. The parameter $\gamma_m$ is the dissipation rate due to the environment and $\xi$ his stochastic effect. We define the noises $\hat{w}_x =\hat{b}_t^{\dagger}+\hat{b}_t$ and $\hat{w}_p = i(\hat{b}_t^{\dagger}-\hat{b}_t)$. We notice that the addition of a collapse-induced dissipative mechanism has changed both the equation for the position and for the momentum. Notably, this raptures the proportionality between the velocity $\D \hat x/\D t$ and the momentum $\hat p$. 
However, from the experimental perspective, the relevant quantity to consider is the second derivative of the position operator, which reads
\begin{equation}
\label{second_deriv_opto_langevin}
\frac{\D^{2} \hat{x}}{\D t^{2}} =-\Omega_0^2 \hat{x} 
- (\Gamma + \gamma_m)\frac{\D\hat{x}}{\D t} + \mathcal{\hat{N}},
\end{equation}
where $\Omega_0^2 = \omega_0^2 + \frac{\Gamma}{2}(\frac{\Gamma}{2}+\gamma_m)$  and $\mathcal{\hat{N}} = -(\frac{\Gamma}{2}+\gamma_m)\hbar\alpha \hat{w}_x+\frac{\xi}{M}-\frac{\hbar\hat{w}_p}{M}-\hbar\alpha\frac{\D \hat{w}_x}{\D t}$. Thus, we have that $\Gamma+\gamma_m$ is the total dissipation rate of the center of mass. This means that the effect of the inclusion of collapse-induced dissipation is to change the total dissipation rate of the mechanical oscillator by a quantity $\Gamma$.

Following the same procedure as that presented in \cite{nobakht2018unitary}, one can derive the corresponding steady-state density noise spectrum \cite{gardiner2004quantum,paternostro2006reconstructing}. By starting from 
 Eq.~\eqref{second_deriv_opto_langevin}, one obtains
\begin{equation}
\label{noise_spectrum}
S(\omega)\!=\!\frac{
\frac{\hbar \omega\gamma_m}{M} \coth(\frac{\hbar\omega}{2\kB T})+\eta\left[\frac{\Gamma^2(2\gamma_m+\Gamma)^2}{16\eta^2}+\frac{\Gamma^2\omega^2}{4\eta^2}+\frac{\hbar^2}{M^2}\right]
}{
\left(\omega_0^2+\frac{\Gamma}{2}\left(\frac{\Gamma}{2}+\gamma_m\right)-\omega^2\right)^2+\omega^2(\Gamma+\gamma_m)^2
},
\end{equation}
where the first term quantifies for the environmental effects, while the second accounts for the collapse-induced ones. Fundamentally, the Lorentzian profile of $S(\omega)$ has a width half-height being equal to $\Gamma+\gamma_m$.

\section{Experimental bounds}
\label{exp_bounds}

\begin{figure}[t]
\centering
\includegraphics[width=0.5\linewidth]{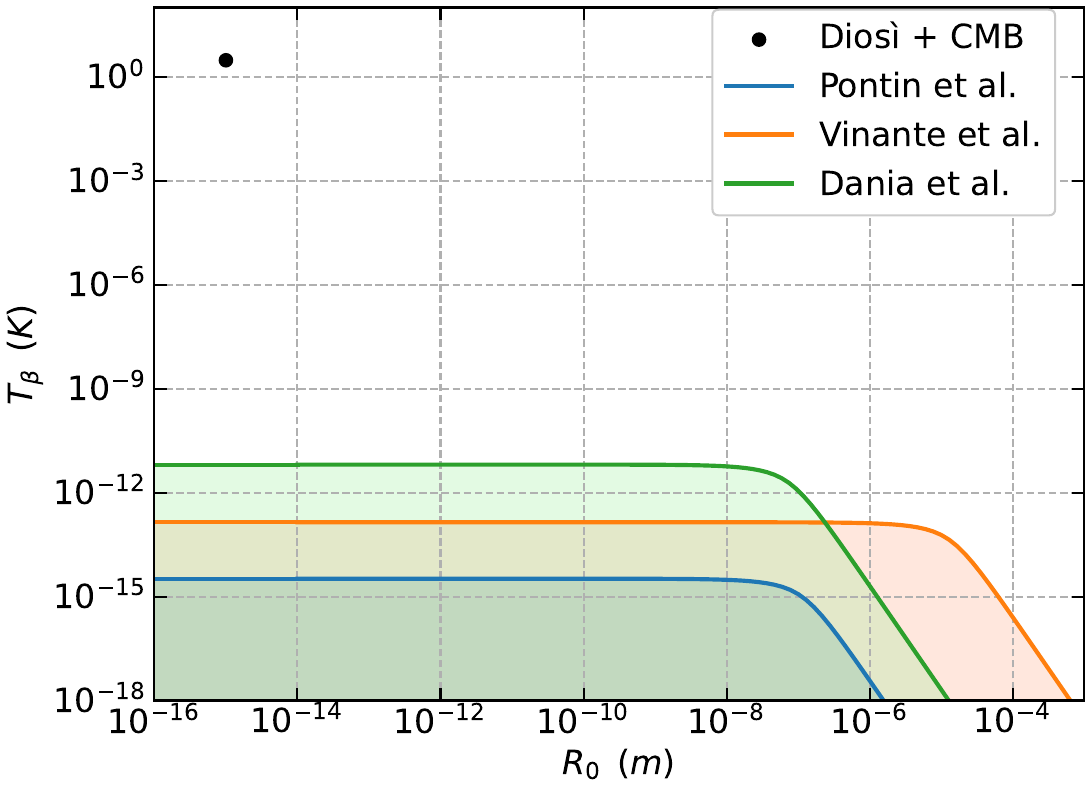}
\caption{{Experimental bounds for the dDP model. } Blue, orange and green shaded areas represent the excluded values of the collapse parameters, respectively, from the experiments of Pontin \textit{et al.},
Vinante \textit{et al.} and Dania \textit{et al.}. The black point represents
the proposal by Diósi of $R_0 = 10^{-15}$\,m and $T_{\beta} \sim 3$\,K being the CMB temperature.}
\label{bounds_dDP}
\end{figure}

\begin{figure}[h]
\centering
\includegraphics[width=0.6\linewidth]{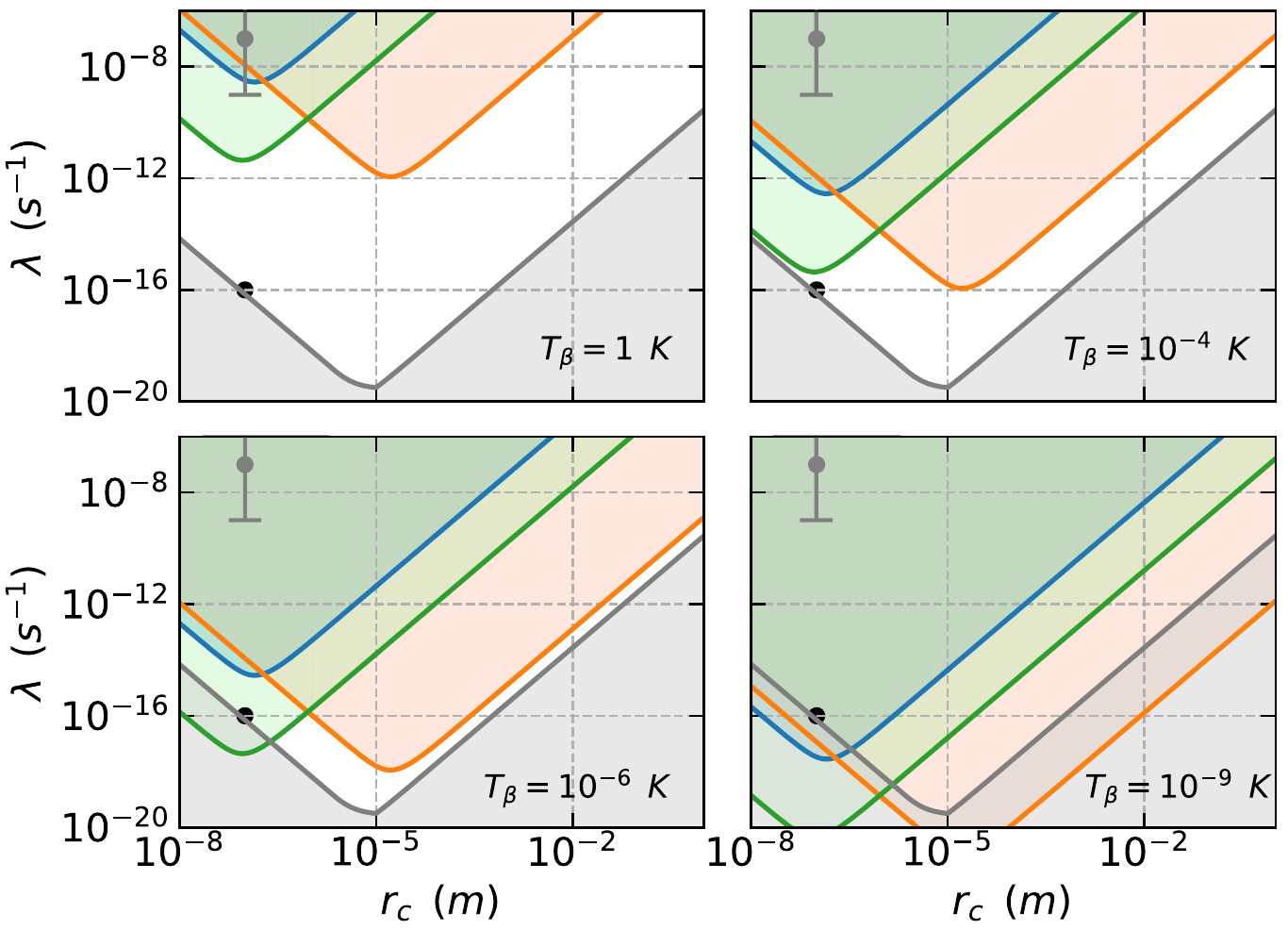}
\caption{{Experimental bounds for the dCSL model.} Blue, orange and green shaded areas represent the excluded values of the collapse parameters, respectively, from the experiments of Pontin \textit{et al.},
Vinante \textit{et al.} and Dania \textit{et al.}. Each panel considers different values of $T_{\beta}$, left-to-right, top-to-bottom, these are 1\,K, $10^{-4}$\,K, $10^{-6}$\,K and $10^{-9}$\,K. The grey area is excluded for theoretical reasons \cite{torovs2017colored}. The grey bar is the Adler proposal of the CSL parameters \cite{adler2007lower}  and the black point is the GRW proposal \cite{ghirardi1986unified}.}
\label{bounds_dCSL}
\end{figure}

Now we are able to set experimental bounds on dDP and dCSL free parameters. We focus on levitated optomechanics, which provides promising platforms for testing fundamental physics and  quantum mechanics \cite{millen2020optomechanics,gonzalez2021levitodynamics,moore2021searching}. In particular, experiments with low dissipation
are of our interest. We notice that it is challenging to establish bounds on the parameter $\Gamma$ from Eq.~\eqref{noise_spectrum} since it appears in different contributions to $S(\omega)$. In contrast, the task becomes straightforward when examining Eq.~\eqref{second_deriv_opto_langevin}.
Indeed, owning the fact that the total dissipation rate in Eq.~\eqref{second_deriv_opto_langevin} is $\Gamma + \gamma_m$, we know that experimentally measured dissipative rate $\gamma_{\text{exp}}$ will provide an estimation of the upper bound for $\Gamma$. Such an estimation is conservative since the value of $\gamma_m$ is fully neglected. Then, by using the expression for $\Gamma$ in Eq.~\eqref{gamma_eta} and defining $T_{\beta} = 1/\kB\beta$ as the temperature of the collapse field, we find
\begin{equation}
    \gamma_\text{exp}\geq\Gamma=\frac{\hbar^2\eta}{2M}\frac{1}{\kB T_\beta},
\end{equation}
where $k_{\text{B}}$ is the Boltzmann constant.
Since for the dDP model, $\eta_{\text{DP}}$ in Eq.~\eqref{etaDP} is a function of  $R_0$ only, we can bound the possible values of $T_\beta$ as a function of $R_0$. Conversely, for the dCSL model $\eta_{\text{CSL}}$ in Eq.~\eqref{etaCSL} is a function of two free parameters ($\lambda$ and $\rC$). Thus, we study how the bounds on $\lambda$ with respect to $\rC$ change when varying the values of $T_\beta$. Specifically, one has

\begin{equation}
T_{\beta} \geq \frac{\hbar^2\eta_{\text{DP}}}{2M k_{\text{B}}\gamma_{\text{exp}}},\quad \lambda \leq \frac{2Mk_{\text{B}}T_{\beta}\gamma_{\text{exp}}}{\hbar^2\bar{\eta}_{\text{CSL}}},
\end{equation}
where $\bar{\eta}_{\text{CSL}} = \eta_{\text{CSL}}/\lambda$. To be quantitative, we use the experimental data from three recent experiments in levitated optomechanics, which are those of Pontin \textit{et al.} \cite{PhysRevResearch.2.023349}, Vinante \textit{et al.} \cite{vinante2020testing} and Dania \textit{et al.} \cite{dania2023ultra}. The first and last experiment use linear Paul traps to levitate a silica nanoparticles of mass and linewidth respectively being
$M= 9.6\times 10^{-17}$\,kg, $\gamma_\text{exp}=2\pi\times48$\,$\mu$Hz and $M=4.3\times 10^{-17}$\,kg,  $\gamma_\text{exp}=2\pi\times 80$\,nHz. Conversely, Vinante uses a lavitated micromagnet of mass $M=6.1\times10^{-10}$\,kg from which one infers a linewidth $\gamma_\text{exp}=2\pi\times9$\,$\mu$Hz at zero pressure.
The experiments of Pontin and Vinante were already used to set bounds on an earlier dissipative version of the DP and CSL model, while that of Dania has not yet been exploited for collapse model testing. 
We show the experimental bounds on the dDP and dCSL respectively in 
Fig.~\ref{bounds_dDP} and Fig.~\ref{bounds_dCSL}. Here, the blue, orange and green shaded areas correspond to the excluded values of the collapse parameters by the  experiments of Pontin \textit{et al.}, Vinante \textit{et al.} and Dania \textit{et al.}

In Fig.~\ref{bounds_dDP} we show the excluded values of $T_\beta$ for the dDP model when varying $R_0$.
For comparison, we also report (black dot) the values of Diósi proposal of $R_0 = 10^{-15}$\,m matched with $T_{\beta} \sim 3$\,K being the temperature of the cosmic microwave background (CMB). This choice is based on the hypothesis of a cosmological origin of the collapse mechanism, and thus one would expect a value of $T_\beta$ of this order of magnitude. We notice that 
below $T_{\beta} \sim 10^{-13}$\,K and $T_{\beta} \sim 6 \times 10^{-12}$\,K all the values of $R_0$ respectively smaller than $10^{-6}$\,m and $10^{-8}$\,m are excluded, this includes the mesoscopic regime where one would expect a collapse.

In Fig.~\ref{bounds_dCSL} we show the bounds on the dCSL parameters $\lambda$ and $\rC$ for four values of $T_{\beta} = 1$\,K, $10^{-3}$\,K, $10^{-5}$\,K and $10^{-7}$\,K. The gray region is excluded theoretically as it would not guarantee an effective collapse of macroscopic quantum superpositions \cite{torovs2017colored}. The grey bar, which is the Adler proposal \cite{adler2007lower} for the CSL parameters, is excluded for each  value of $T_\beta$ reported, while the GRW proposal \cite{ghirardi1986unified} is excluded for $T_\beta=10^{-5}$\,K and below. We notice that all the parameter space is excluded for temperatures lower than $ 6 \times 10^{-9}$\,K . 

\section{Comparison with the previous dissipative model}\label{comparison}
Here we compare the linear friction (LF) dissipative model, introduced in Ref.~\cite{PhysRevA.108.012202} and shown in Eq.~\eqref{ME}, with the previously proposed dissipative collapse models \cite{smirne2015dissipative,bahrami2014role}. The latter have a mathematical structure similar to the collisional dynamics of a test particle interacting with a low-density gas in the weak coupling regime \cite{vacchini2000completely}. Thus, for simplicity, we refer to these as collisional dynamics (CD) models.  For such a comparison, we compute the asymptotic temperature of the center of mass of the mechanical oscillator, which in both frameworks can be derived from their respective master equations. Indeed, given an 
arbitrary operator $\hat{O}$, one can compute the evolution of its expectation value as $\frac{\D}{\D t}\braket{\hat{O}}_t=\Tr(\hat{O}\frac{\D}{\D t}\hat{\rho}_{cm})$. 
The equation with $\hat{O}=\hat{H}$ is not in a closed form, however we can write the system of three differential equations for $\hat{O}=\hat{V}$, $\hat{K}$ and $\{\hat{x},\hat{p}\}$, where $\hat{K} = \hat{p}^2/2M$ 
and $\hat{V}=M\omega_0^2\hat{x}^2/2$. Under the assumption of a reaching a stable condition at the thermal equilibrium, we can set all the derivatives to zero and find the asymptotic values of $\braket{\hat{K}}_{\infty}$ and $\braket{\hat{V}}_{\infty}$, from which we obtain $\braket{\hat{H}}_{\infty} = \braket{\hat{K}}_{\infty}+\braket{\hat{V}}_{\infty}$. Then, we define the temperature of the system by exploiting the equipartition theorem for single harmonic oscillator $\braket{\hat{H}}_{\infty} = \kB T$.

In the LF framework, we have
\begin{equation}\label{LF.system}
    \begin{aligned}
        \frac{\D \braket{\hat{V}}_t}{\D t}&=-\Gamma \braket{\hat{V}}_t+\frac{\omega_0^2}{2}\braket{\{\hat{x},\hat{p}\}}_t+\frac{\Gamma^2 M \omega_0^2}{8\eta},\\
        \frac{\D \braket{\hat{K}}_t}{\D t}&=-\Gamma \braket{\hat{K}}_t-\frac{\omega_0^2}{2}\braket{\{\hat{x},\hat{p}\}}_t+\frac{\hbar^2 \eta}{2M},\\
        \frac{\D \braket{\{\hat x,\hat p\}}_t}{\D t}&=-\Gamma \braket{\{\hat{x},\hat{p}\}}_t+4\braket{\hat{K}}_t-4\braket{\hat{V}}_t,\\
    \end{aligned}
\end{equation}
\begin{figure*}[t]
\centering
\includegraphics[width=0.48\linewidth]{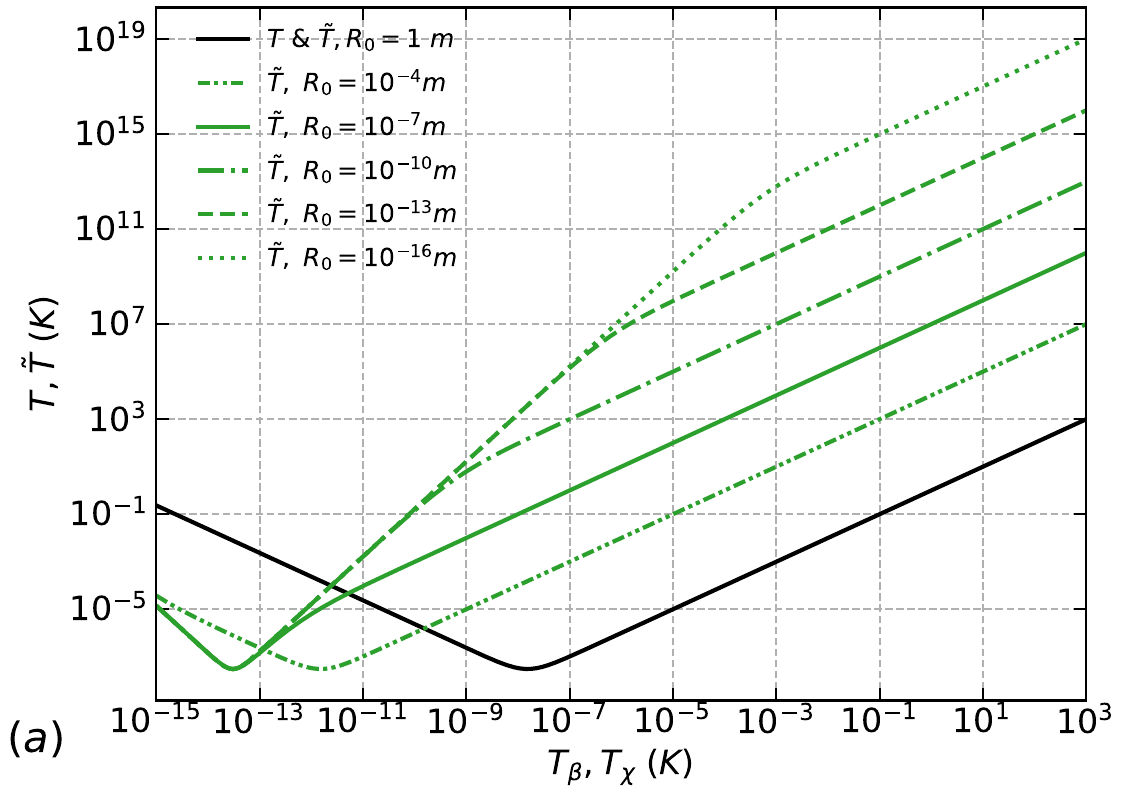} \includegraphics[width=0.48\linewidth]{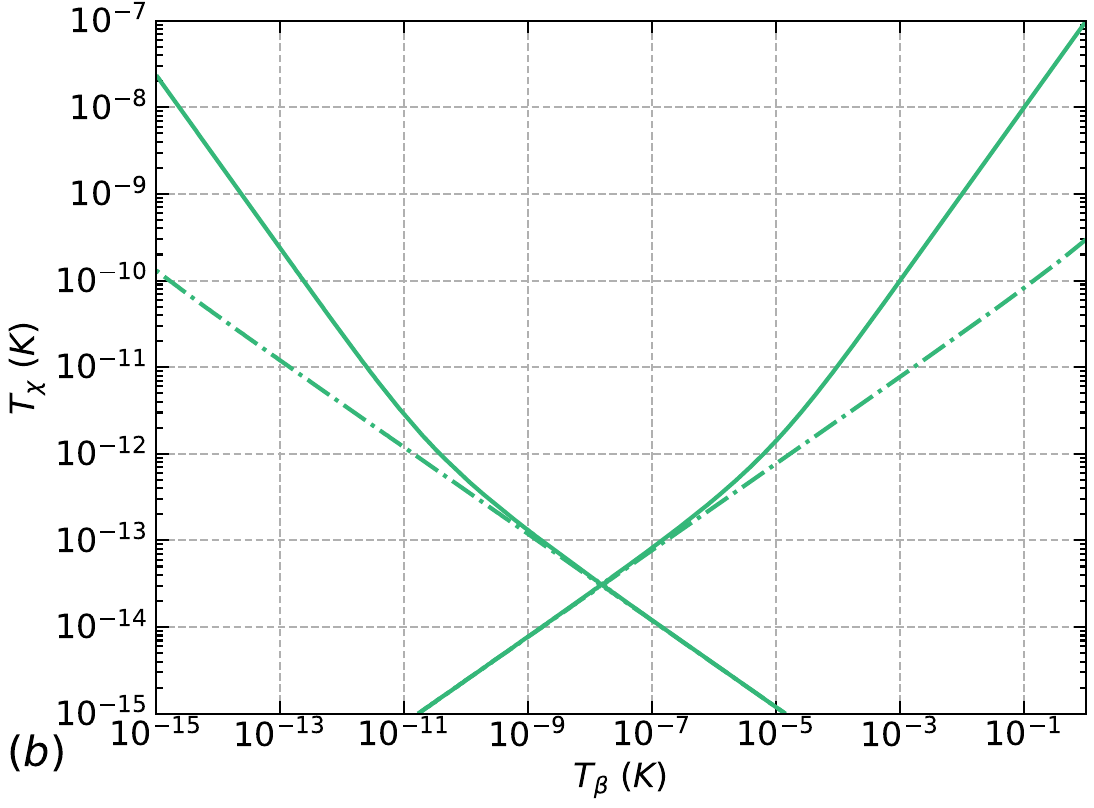}
\caption{{Comparison between the linear friction dDP and the collisional dynamics dDP models.} In panel (a) the solid black curve represents the asymptotic temperature $T$ of the LF-dDP model as a function of the dissipation parameter $T_{\beta}$, while the green curves show the asymptotic temperature $\tilde{T}$ of the CD-dDP model as a function of the dissipation parameter $T_{\chi}$ for different values of $R_0$. For $R_0 = 1$ m, $T$ and $\tilde{T}$ coincide in the black curve. Panel (b) shows the contour plot of the function $T(T_{\beta}) - \tilde{T}(T_{\chi}) = 0$ where the solid green line is for $R_{0} = 10^{-7}$\,m and the dash dotted green line is for $R_{0} = 10^{-10}$\,m.}
\label{comp_DP}
\end{figure*}
which correspond to 
\begin{equation}
\label{T_asymptotic}
T = T_{\beta} + \frac{\hbar^2\omega_0^2}{16\kB^2 T_{\beta}},
\end{equation}
both for the dDP and the dCSL model.
We notice that Eq.~\eqref{T_asymptotic} does not depend on the free parameters of the model except for the dissipation parameter $\beta=(\kB T_\beta)^{-1}$. When $T_{\beta}$ is high, the asymptotic temperature $T$ coincides with the collapse temperature $T_{\beta}$. Indeed, in the limit of $T_{\beta}\to\infty$ ($\beta\to0$, i.e.~$\Gamma\to0$), the last term of the first expression in Eq.~\eqref{LF.system} can be neglected and the only important collapse term is the last one in the second expression. In such a way
one recovers the predictions of the standard collapse model without dissipation, for which one has $T = \infty$. 
Also in the limit of $T_{\beta}\to 0$, the asymptotic temperature $T$ goes to infinity. Indeed, in such a limit, is the last term in the second expression of Eq.~\eqref{LF.system} that can be neglected, and the last term in the first expression becomes the relevant one. The latter leads to an infinite increase to the mean potential energy, and thus to $\braket{\hat{H}}_{\infty}=\infty$.

In the CD framework, one has
\begin{equation}\label{CD.system}
    \begin{aligned}
        \frac{\D \braket{\hat{V}}_t}{\D t}&=\frac{\omega_0^2}{2}\braket{\{\hat{x},\hat{p}\}}_t+\frac{\tilde{\Gamma}^2 M \omega_0^2}{8 \tilde{\eta}},\\
        \frac{\D \braket{\hat{K}}_t}{\D t}&=-2\tilde{\Gamma}\braket{\hat{K}}_t-\frac{\omega_0^2}{2}\braket{\{\hat{x},\hat{p}\}}_t+\frac{\hbar^2 \tilde{\eta}}{2M},\\
        \frac{\D \braket{\{\hat x,\hat p\}}_t}{\D t}&=-\tilde{\Gamma} \braket{\{\hat{x},\hat{p}\}}_t+4\braket{\hat{K}}_t-4\braket{\hat{V}}_t,\\
    \end{aligned}
\end{equation}
whose corresponding  asymptotic temperature reads
\begin{equation}
\tilde{T} = \frac{\tilde{\Gamma} M\omega_0^2}{8\tilde{\eta}\kB}+\frac{\hbar^2\tilde{\eta}}{2M\tilde{\Gamma}\kB}+\frac{\tilde{\Gamma}^3 M}{16 \tilde{\eta}\kB}.
\end{equation}
\begin{figure*}[t!]
\centering
\includegraphics[width=0.48\linewidth]{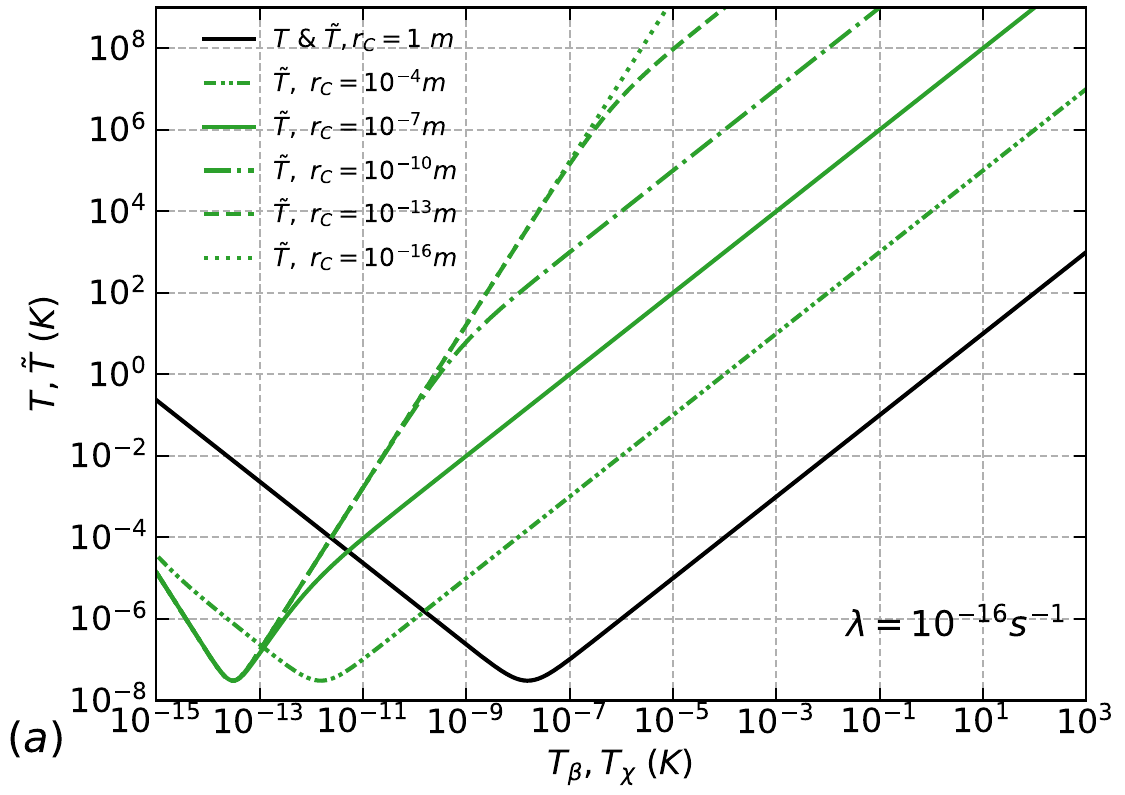} \includegraphics[width=0.48\linewidth]{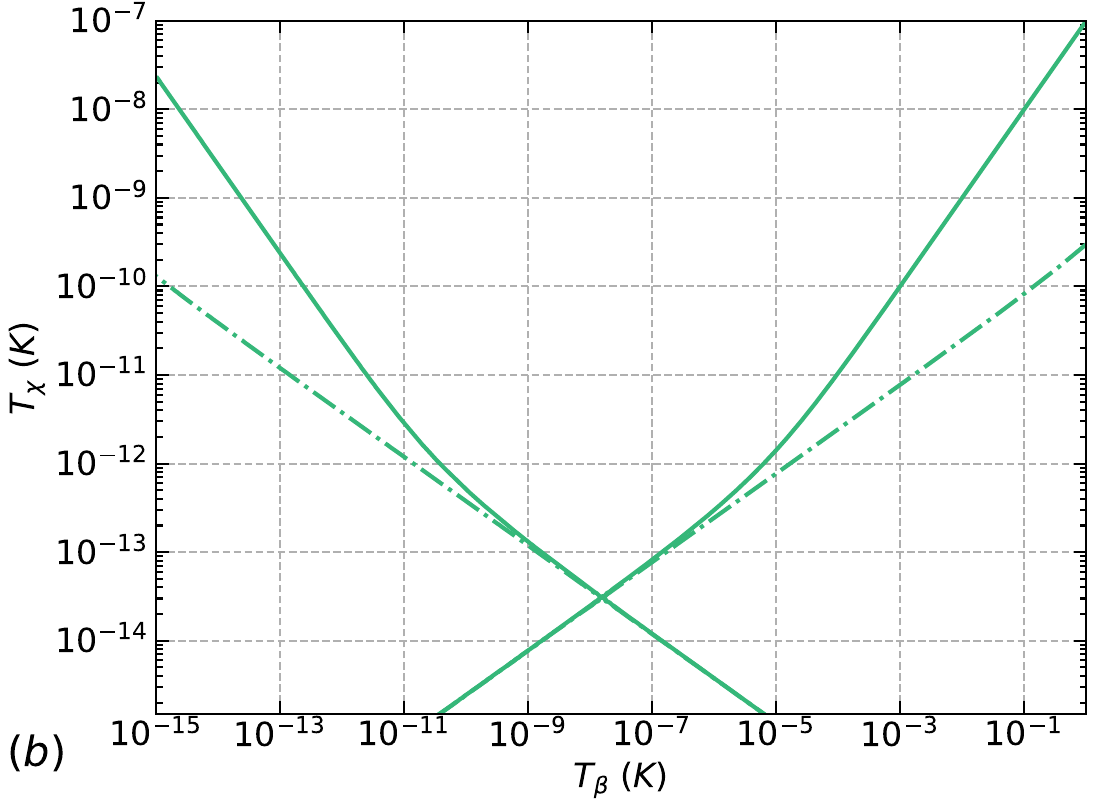}
\caption{{Comparison between the linear friction dCSL and the collisional dynamics dCSL models.} In panel (a) the solid black curve represents the asymptotic temperature $T$ of the LF-dCSL model as a function of the dissipation parameter $T_{\beta}$, while the green curves show the asymptotic temperature $\tilde{T}$ of the CD-dCSL model as a function of the dissipation parameter $T_{\chi}$ for different values of $\rC$. For $\rC = 1$ m, $T$ and $\tilde{T}$ coincide in the black curve. Panel (b) shows the contour plot of the function $T(T_{\beta}) - \tilde{T}(T_{\chi}) = 0$ where the solid green line is for $\rC = 10^{-7}$\,m and the dash dotted green line is for $\rC = 10^{-10}$\,m. We fix $\lambda = 10^{-16}$\,s$^{-1}$.}
\label{comp_CSL}
\end{figure*}
The latter depends on two parameters $\tilde \eta$ and $\tilde \Gamma$ that play the role, respectively, of $\eta$ and $\Gamma$ in the LF framework \cite{nobakht2018unitary}. To be specific, we have  $\tilde{\Gamma} = 4\tilde{\eta}\sigma^2\chi(1+\chi)m_0/M$ with $\sigma = R_0$ or $\rC$ and where $\chi = \hbar^2/8m_0\sigma^2 \kB T_{\chi}$ is the dissipation parameter of the CD model and $T_{\chi}$ is the associated temperature of the collapse field (analogous to $T_\beta$ in the LF framework). The coefficient $\tilde{\eta}$ takes the following form 
\begin{equation}
\label{eta_tildeDP}
\begin{aligned}
&\tilde{\eta}_\text{DP} = \sqrt{\pi}  \operatorname{erf}\left(\frac{r}{\tilde R_0 }\right)
+\frac{G M^2R_0}{\sqrt{\pi}r^3}\Biggl[\frac{ R_1 }{r}
\left(e^{-\frac{r^2}{ R_1^2}}-3\right)+2\frac{ R_1^3}{r^3}\left(1-e^{-\frac{r^2}{ R_1^2}}\right)
\Biggr],
\end{aligned}
\end{equation}
for the CD-dDP model with $R_1=R_0(1+\chi)$, and 
\begin{equation}
\label{eta_tildeCSL}
\tilde{\eta}_\text{CSL} = \frac{3\lambda M^2 \rC^3}{\RC m_0^2 r^4}\Biggl[1-2\left(\frac{\RC }{r}\right)^2+e^{-\frac{r^2}{\RC^2 }}\left(1+2\frac{\RC^2 }{r^2}\right)\Biggr],
\end{equation}
for the CD-dCSL model with $\RC=\rC(1+\chi)$.
Notably, in the CD framework, $\tilde{T}$ depends on all the free parameters of the CD model. In the limit $T_{\chi} \to \infty$ ($\chi \to 0$), one recovers the standard collapse model with $\tilde{T} = \infty$. In the opposite limit, for $T_{\chi} \to 0$ (i.e.~$\tilde\Gamma\to\infty$), the last term of the first expression of Eq.~\eqref{CD.system} is the relevant one, while the last of the second expression can be neglected. Then, following the same reasoning as in the LF framework, one has $\tilde T=\infty$. Table \ref{table} presents a direct comparison between the parameters  of the two models.

In Fig.~\ref{comp_DP} we compare LF-dDP and CD-dDP models, where the experimental values considered are the mass and the radius of the nano-particle from Dania \textit{et al.} \cite{dania2023ultra}. In panel (a) we show in black the plot of $T$ as a function of $T_{\beta}$ and in green the plots of $\tilde{T}$ as a function of $T_{\chi}$ for various values of $R_0$. We notice that $T$ and $\tilde{T}$ coincide for $R_0 = 1$\,m. As $R_0$ decreases, the difference between $T$ and $\tilde{T}$ increases.

More interestingly if we assume that both the models reach the same asymptotic temperature, namely  $T = \tilde{T}$, then we can link the two dissipation parameters $T_{\beta}$ and $T_{\chi}$ and display how they are related. Thus, in panel (b) of Fig.~\ref{comp_DP} we show the  plot of the function $T(T_{\beta}) - \tilde{T}(T_{\chi}) = 0$. The solid green line is for $R_0 = 10^{-7}$\,m and the dash dotted one for $R_0 = 10^{-10}$\,m. In general the relation between $T_{\beta}$ and $T_{\chi}$ is non-linear and it does not lead to a one-to-one relation. However, in some regimes, we have a linear behaviour and we can compare 
the two collapse temperatures $T_\beta$ and $T_\chi$ directly. For example $T_{\beta} = 1$\,K corresponds to $T_{\chi} = 10^{-7}$\,K for $R_0 = 10^{-7}$\,m (solid line) and to $T_{\chi} \sim 10^{-10}$\,K for $R_0 = 10^{-10}$\,m (dashed line).
\renewcommand{\arraystretch}{2.2}
\begin{table}[b!]
\vspace{5mm}
\centering
\begin{tabular}{|c|c|}
\hline
Linear Friction (LF) model & Collisional Dynamics (CD) model \\
\hline
$\hat{L}_\kv=\rh{\mu}_\kv +\frac{\hbar\beta}{4}\kv \J_\kv$ & $\hat{\tilde{L}}_\kv = m \sum_{j=1}^{N}e^{i\kv\hat{\xv}_{j}}e^{-2\sigma^2[(1+\chi)\kv \hat{\pvec}_j+2k^2\hat{\pvec}_j^2]}$\\
\hline 
%\midrule
$D_\kv =\exp(-\sigma^2 k^2)\times
\begin{cases}
\hbar^2\gamma & (\text{CSL})\\
4\pi\hbar G/k^2 & (\text{DP})
\end{cases}$&$\tilde{D}_\kv =\exp\bigl(-\sigma^2 k^2(1+\chi)^2\bigr)\times
\begin{cases}
\hbar^2\gamma & (\text{CSL})\\
4\pi\hbar G/k^2 & (\text{DP})
\end{cases}$\\
\hline
$\Gamma = \frac{\hbar^{2}\eta}{2M \kB T_{\beta}}$& $\tilde{\Gamma} = \frac{\hbar^2\tilde{\eta}}{2 M \kB T_{\chi}}\left(1+\frac{\hbar^2}{8m_0\sigma^2 \kB T_{\chi}}\right)$\\
\hline
$\eta = \frac{1}{\hbar^{2}}\int \frac{\D^{3}k}{(2\pi)^3}k_{x}^{2}D_{\kv}|\mu_{\kv}|^{2}$ & $\tilde{\eta} = \frac{1}{\hbar^{2}}\int \frac{\D^{3}k}{(2\pi)^3}k_{x}^{2}\tilde{D}_{\kv}|\mu_{\kv}|^{2}$ \\
\hline
$T = T_{\beta} + \frac{\hbar^2\omega_0^2}{16\kB^2 T_{\beta}}$  & $\tilde{T} = \frac{{T}_{\chi}}{1+\frac{\hbar^2}{8m_0\sigma^2\kB T_{\chi}}} + \frac{\hbar^2\omega_0^2}{16\kB^2 T_{\chi}}\left(1+\frac{\hbar^2}{8m_0\sigma^2\kB T_{\chi}}\right) + \frac{\hbar^6\tilde{\eta}^2}{128M^2\kB^4 T_{\chi}^3}\left(1+\frac{\hbar^2}{8m_0\sigma^2\kB T_{\chi}}\right)^3$  \\
\hline
%\bottomrule
\end{tabular}
\caption{Comparison between the parameters of the Linear Friction and Collisional Dynamics dissipative models. Here $\sigma = R_0$, $\rC$ for DP and CSL respectively, $\gamma = (\sqrt{4\pi}\sigma)^3 \lambda / m_0^2$ and $\chi = \hbar^2/8m_0\sigma^2 \kB T_{\chi}$. $\hat{\mu}_{\kv}$ and $\J_\kv$ are defined in Eq.~\eqref{mass_current_fourier}.}
\label{table}
\end{table}
We show the same analysis for LF-dCSL and CD-dCSL models in Fig.~\ref{comp_CSL} where we used the same colouring and dashing as in Fig.~\ref{comp_DP}, and where we set $\lambda = 10^{-16}$\,s$^{-1}$ corresponding to the GRW point at $\rC=10^{-7}$\,m. We notice that the LF-dCSL and CD-dCSL models lead two different predictions. This is exemplified by the GRW point, which in the CD-dCSL model is excluded for collapse temperatures $T_\chi$ below $10^{-9}$\,K (see Ref.~\cite{vinante2020testing}. On the other hand, focusing on the top right branch of the solid line  in Fig.~\ref{comp_CSL}b, the value of $T_\chi=10^{-9}$\,K corresponds to $T_\beta=10^{-2}$\,K for which the GRW point is not excluded [cf.~Fig.~\ref{bounds_dCSL}].
This means that the two frameworks, LF and CD, can be in principle discriminates experimentally. A similar example can be showcased in the comparison of the LF-dDP and CD-dDP models. Notably, the relation between $T_\chi$ and $T_\beta$ for the dDP and dCSL models show the same behaviour [cf.~Fig.~\ref{comp_DP}b and Fig.~\ref{comp_CSL}b].

\section{Conclusions and Outlook}\label{conclusion}
A new mechanism to introduce dissipation in collapse models has been recently proposed. Conversely to a previously proposed one (indicated as Collisional Dynamics (CD) dissipative model), this mechanism is based on the linear-friction of the current being linear in the current of the many-body system (thus, named as Linear Friction (LF) dissipative model). Due to this feature, the LF model has a more physical appeal than the CD model. In addition, LF is easier to investigate, as evidenced in Table \ref{table}, which enables a comparison of the parameters of the two models.
LF model has not yet been tested, opens a promising avenue for new investigations in the collapse models framework. We focus on establishing the first experimental bounds for linear-friction dissipative DP (dDP) and CSL (dCSL) models, using data from levitated optomechanical experiments. The results reveal significant exclusions of the parameter space, with collapse temperatures below $T_\beta\sim 10^{-13}$\,K and $T_\beta\sim 6\times 10^{-12}$\,K for dDP model and all parameter space for dCSL model is excluded for temperatures below $T_\beta\sim 6 \times 10^{-9}$\,K. Finally, we compare the two models. We find the relations between the respective collapse temperature under the assumption that the collapse process leads the system to thermalisation. We conclude that they can in principle be discriminated experimentally. 

\section*{Acknowledgments}
The authors thank useful discussion with Lajos Diosi. We acknowledge the University of Trieste, INFN, EIC Pathfinder project QuCoM (GA No.~101046973), and the PNRR PE National Quantum Science and Technology Institute (PE0000023).

\bibliography{biblio}

%apsrev4-2.bst 2019-01-14 (MD) hand-edited version of apsrev4-1.bst
%Control: key (0)
%Control: author (72) initials jnrlst
%Control: editor formatted (1) identically to author
%Control: production of article title (-1) disabled
%Control: page (0) single
%Control: year (1) truncated
%Control: production of eprint (0) enabled
\begin{thebibliography}{37}%
\makeatletter
\providecommand \@ifxundefined [1]{%
 \@ifx{#1\undefined}
}%
\providecommand \@ifnum [1]{%
 \ifnum #1\expandafter \@firstoftwo
 \else \expandafter \@secondoftwo
 \fi
}%
\providecommand \@ifx [1]{%
 \ifx #1\expandafter \@firstoftwo
 \else \expandafter \@secondoftwo
 \fi
}%
\providecommand \natexlab [1]{#1}%
\providecommand \enquote  [1]{``#1''}%
\providecommand \bibnamefont  [1]{#1}%
\providecommand \bibfnamefont [1]{#1}%
\providecommand \citenamefont [1]{#1}%
\providecommand \href@noop [0]{\@secondoftwo}%
\providecommand \href [0]{\begingroup \@sanitize@url \@href}%
\providecommand \@href[1]{\@@startlink{#1}\@@href}%
\providecommand \@@href[1]{\endgroup#1\@@endlink}%
\providecommand \@sanitize@url [0]{\catcode `\\12\catcode `\$12\catcode `\&12\catcode `\#12\catcode `\^12\catcode `\_12\catcode `\%12\relax}%
\providecommand \@@startlink[1]{}%
\providecommand \@@endlink[0]{}%
\providecommand \url  [0]{\begingroup\@sanitize@url \@url }%
\providecommand \@url [1]{\endgroup\@href {#1}{\urlprefix }}%
\providecommand \urlprefix  [0]{URL }%
\providecommand \Eprint [0]{\href }%
\providecommand \doibase [0]{https://doi.org/}%
\providecommand \selectlanguage [0]{\@gobble}%
\providecommand \bibinfo  [0]{\@secondoftwo}%
\providecommand \bibfield  [0]{\@secondoftwo}%
\providecommand \translation [1]{[#1]}%
\providecommand \BibitemOpen [0]{}%
\providecommand \bibitemStop [0]{}%
\providecommand \bibitemNoStop [0]{.\EOS\space}%
\providecommand \EOS [0]{\spacefactor3000\relax}%
\providecommand \BibitemShut  [1]{\csname bibitem#1\endcsname}%
\let\auto@bib@innerbib\@empty
%</preamble>
\bibitem [{\citenamefont {Bassi}\ and\ \citenamefont {Ghirardi}(2003)}]{bassi2003dynamical}%
  \BibitemOpen
  \bibfield  {author} {\bibinfo {author} {\bibfnamefont {A.}~\bibnamefont {Bassi}}\ and\ \bibinfo {author} {\bibfnamefont {G.}~\bibnamefont {Ghirardi}},\ }\href@noop {} {\bibfield  {journal} {\bibinfo  {journal} {Physics Reports}\ }\textbf {\bibinfo {volume} {379}},\ \bibinfo {pages} {257} (\bibinfo {year} {2003})}\BibitemShut {NoStop}%
\bibitem [{\citenamefont {Bassi}\ \emph {et~al.}(2013)\citenamefont {Bassi}, \citenamefont {Lochan}, \citenamefont {Satin}, \citenamefont {Singh},\ and\ \citenamefont {Ulbricht}}]{bassi2013models}%
  \BibitemOpen
  \bibfield  {author} {\bibinfo {author} {\bibfnamefont {A.}~\bibnamefont {Bassi}}, \bibinfo {author} {\bibfnamefont {K.}~\bibnamefont {Lochan}}, \bibinfo {author} {\bibfnamefont {S.}~\bibnamefont {Satin}}, \bibinfo {author} {\bibfnamefont {T.~P.}\ \bibnamefont {Singh}},\ and\ \bibinfo {author} {\bibfnamefont {H.}~\bibnamefont {Ulbricht}},\ }\href@noop {} {\bibfield  {journal} {\bibinfo  {journal} {Reviews of Modern Physics}\ }\textbf {\bibinfo {volume} {85}},\ \bibinfo {pages} {471} (\bibinfo {year} {2013})}\BibitemShut {NoStop}%
\bibitem [{\citenamefont {Carlesso}\ \emph {et~al.}(2022)\citenamefont {Carlesso}, \citenamefont {Donadi}, \citenamefont {Ferialdi}, \citenamefont {Paternostro}, \citenamefont {Ulbricht},\ and\ \citenamefont {Bassi}}]{carlesso2022present}%
  \BibitemOpen
  \bibfield  {author} {\bibinfo {author} {\bibfnamefont {M.}~\bibnamefont {Carlesso}}, \bibinfo {author} {\bibfnamefont {S.}~\bibnamefont {Donadi}}, \bibinfo {author} {\bibfnamefont {L.}~\bibnamefont {Ferialdi}}, \bibinfo {author} {\bibfnamefont {M.}~\bibnamefont {Paternostro}}, \bibinfo {author} {\bibfnamefont {H.}~\bibnamefont {Ulbricht}},\ and\ \bibinfo {author} {\bibfnamefont {A.}~\bibnamefont {Bassi}},\ }\href@noop {} {\bibfield  {journal} {\bibinfo  {journal} {Nature Physics}\ }\textbf {\bibinfo {volume} {18}},\ \bibinfo {pages} {243} (\bibinfo {year} {2022})}\BibitemShut {NoStop}%
\bibitem [{\citenamefont {Bilardello}\ \emph {et~al.}(2016)\citenamefont {Bilardello}, \citenamefont {Donadi}, \citenamefont {Vinante},\ and\ \citenamefont {Bassi}}]{Bilardello:2016aa}%
  \BibitemOpen
  \bibfield  {author} {\bibinfo {author} {\bibfnamefont {M.}~\bibnamefont {Bilardello}}, \bibinfo {author} {\bibfnamefont {S.}~\bibnamefont {Donadi}}, \bibinfo {author} {\bibfnamefont {A.}~\bibnamefont {Vinante}},\ and\ \bibinfo {author} {\bibfnamefont {A.}~\bibnamefont {Bassi}},\ }\href@noop {} {\bibfield  {journal} {\bibinfo  {journal} {Physica A}\ }\textbf {\bibinfo {volume} {462}},\ \bibinfo {pages} {764 } (\bibinfo {year} {2016})}\BibitemShut {NoStop}%
\bibitem [{\citenamefont {Vinante}\ \emph {et~al.}(2016)\citenamefont {Vinante}, \citenamefont {Bahrami}, \citenamefont {Bassi}, \citenamefont {Usenko}, \citenamefont {Wijts},\ and\ \citenamefont {Oosterkamp}}]{vinante2016upper}%
  \BibitemOpen
  \bibfield  {author} {\bibinfo {author} {\bibfnamefont {A.}~\bibnamefont {Vinante}}, \bibinfo {author} {\bibfnamefont {M.}~\bibnamefont {Bahrami}}, \bibinfo {author} {\bibfnamefont {A.}~\bibnamefont {Bassi}}, \bibinfo {author} {\bibfnamefont {O.}~\bibnamefont {Usenko}}, \bibinfo {author} {\bibfnamefont {G.}~\bibnamefont {Wijts}},\ and\ \bibinfo {author} {\bibfnamefont {T.}~\bibnamefont {Oosterkamp}},\ }\href@noop {} {\bibfield  {journal} {\bibinfo  {journal} {Physical review letters}\ }\textbf {\bibinfo {volume} {116}},\ \bibinfo {pages} {090402} (\bibinfo {year} {2016})}\BibitemShut {NoStop}%
\bibitem [{\citenamefont {Carlesso}\ \emph {et~al.}(2016)\citenamefont {Carlesso}, \citenamefont {Bassi}, \citenamefont {Falferi},\ and\ \citenamefont {Vinante}}]{PhysRevD.94.124036}%
  \BibitemOpen
  \bibfield  {author} {\bibinfo {author} {\bibfnamefont {M.}~\bibnamefont {Carlesso}}, \bibinfo {author} {\bibfnamefont {A.}~\bibnamefont {Bassi}}, \bibinfo {author} {\bibfnamefont {P.}~\bibnamefont {Falferi}},\ and\ \bibinfo {author} {\bibfnamefont {A.}~\bibnamefont {Vinante}},\ }\href@noop {} {\bibfield  {journal} {\bibinfo  {journal} {Phys. Rev. D}\ }\textbf {\bibinfo {volume} {94}},\ \bibinfo {pages} {124036} (\bibinfo {year} {2016})}\BibitemShut {NoStop}%
\bibitem [{\citenamefont {Adler}\ \emph {et~al.}(2019)\citenamefont {Adler}, \citenamefont {Bassi}, \citenamefont {Carlesso},\ and\ \citenamefont {Vinante}}]{PhysRevD.99.103001}%
  \BibitemOpen
  \bibfield  {author} {\bibinfo {author} {\bibfnamefont {S.~L.}\ \bibnamefont {Adler}}, \bibinfo {author} {\bibfnamefont {A.}~\bibnamefont {Bassi}}, \bibinfo {author} {\bibfnamefont {M.}~\bibnamefont {Carlesso}},\ and\ \bibinfo {author} {\bibfnamefont {A.}~\bibnamefont {Vinante}},\ }\href@noop {} {\bibfield  {journal} {\bibinfo  {journal} {Phys. Rev. D}\ }\textbf {\bibinfo {volume} {99}},\ \bibinfo {pages} {103001} (\bibinfo {year} {2019})}\BibitemShut {NoStop}%
\bibitem [{\citenamefont {Vinante}\ \emph {et~al.}(2017)\citenamefont {Vinante}, \citenamefont {Mezzena}, \citenamefont {Falferi}, \citenamefont {Carlesso},\ and\ \citenamefont {Bassi}}]{vinante2017improved}%
  \BibitemOpen
  \bibfield  {author} {\bibinfo {author} {\bibfnamefont {A.}~\bibnamefont {Vinante}}, \bibinfo {author} {\bibfnamefont {R.}~\bibnamefont {Mezzena}}, \bibinfo {author} {\bibfnamefont {P.}~\bibnamefont {Falferi}}, \bibinfo {author} {\bibfnamefont {M.}~\bibnamefont {Carlesso}},\ and\ \bibinfo {author} {\bibfnamefont {A.}~\bibnamefont {Bassi}},\ }\href@noop {} {\bibfield  {journal} {\bibinfo  {journal} {Physical review letters}\ }\textbf {\bibinfo {volume} {119}},\ \bibinfo {pages} {110401} (\bibinfo {year} {2017})}\BibitemShut {NoStop}%
\bibitem [{\citenamefont {Helou}\ \emph {et~al.}(2017)\citenamefont {Helou}, \citenamefont {Slagmolen}, \citenamefont {McClelland},\ and\ \citenamefont {Chen}}]{helou2017lisa}%
  \BibitemOpen
  \bibfield  {author} {\bibinfo {author} {\bibfnamefont {B.}~\bibnamefont {Helou}}, \bibinfo {author} {\bibfnamefont {B.}~\bibnamefont {Slagmolen}}, \bibinfo {author} {\bibfnamefont {D.~E.}\ \bibnamefont {McClelland}},\ and\ \bibinfo {author} {\bibfnamefont {Y.}~\bibnamefont {Chen}},\ }\href@noop {} {\bibfield  {journal} {\bibinfo  {journal} {Physical Review D}\ }\textbf {\bibinfo {volume} {95}},\ \bibinfo {pages} {084054} (\bibinfo {year} {2017})}\BibitemShut {NoStop}%
\bibitem [{\citenamefont {Vinante}\ \emph {et~al.}(2020{\natexlab{a}})\citenamefont {Vinante}, \citenamefont {Carlesso}, \citenamefont {Bassi}, \citenamefont {Chiasera}, \citenamefont {Varas}, \citenamefont {Falferi}, \citenamefont {Margesin}, \citenamefont {Mezzena},\ and\ \citenamefont {Ulbricht}}]{PhysRevLett.125.100404}%
  \BibitemOpen
  \bibfield  {author} {\bibinfo {author} {\bibfnamefont {A.}~\bibnamefont {Vinante}}, \bibinfo {author} {\bibfnamefont {M.}~\bibnamefont {Carlesso}}, \bibinfo {author} {\bibfnamefont {A.}~\bibnamefont {Bassi}}, \bibinfo {author} {\bibfnamefont {A.}~\bibnamefont {Chiasera}}, \bibinfo {author} {\bibfnamefont {S.}~\bibnamefont {Varas}}, \bibinfo {author} {\bibfnamefont {P.}~\bibnamefont {Falferi}}, \bibinfo {author} {\bibfnamefont {B.}~\bibnamefont {Margesin}}, \bibinfo {author} {\bibfnamefont {R.}~\bibnamefont {Mezzena}},\ and\ \bibinfo {author} {\bibfnamefont {H.}~\bibnamefont {Ulbricht}},\ }\href@noop {} {\bibfield  {journal} {\bibinfo  {journal} {Phys. Rev. Lett.}\ }\textbf {\bibinfo {volume} {125}},\ \bibinfo {pages} {100404} (\bibinfo {year} {2020}{\natexlab{a}})}\BibitemShut {NoStop}%
\bibitem [{\citenamefont {Zheng}\ \emph {et~al.}(2020)\citenamefont {Zheng}, \citenamefont {Leng}, \citenamefont {Kong}, \citenamefont {Li}, \citenamefont {Wang}, \citenamefont {Luo}, \citenamefont {Zhao}, \citenamefont {Duan}, \citenamefont {Huang}, \citenamefont {Du}, \citenamefont {Carlesso},\ and\ \citenamefont {Bassi}}]{PhysRevResearch.2.013057}%
  \BibitemOpen
  \bibfield  {author} {\bibinfo {author} {\bibfnamefont {D.}~\bibnamefont {Zheng}}, \bibinfo {author} {\bibfnamefont {Y.}~\bibnamefont {Leng}}, \bibinfo {author} {\bibfnamefont {X.}~\bibnamefont {Kong}}, \bibinfo {author} {\bibfnamefont {R.}~\bibnamefont {Li}}, \bibinfo {author} {\bibfnamefont {Z.}~\bibnamefont {Wang}}, \bibinfo {author} {\bibfnamefont {X.}~\bibnamefont {Luo}}, \bibinfo {author} {\bibfnamefont {J.}~\bibnamefont {Zhao}}, \bibinfo {author} {\bibfnamefont {C.-K.}\ \bibnamefont {Duan}}, \bibinfo {author} {\bibfnamefont {P.}~\bibnamefont {Huang}}, \bibinfo {author} {\bibfnamefont {J.}~\bibnamefont {Du}}, \bibinfo {author} {\bibfnamefont {M.}~\bibnamefont {Carlesso}},\ and\ \bibinfo {author} {\bibfnamefont {A.}~\bibnamefont {Bassi}},\ }\href@noop {} {\bibfield  {journal} {\bibinfo  {journal} {Phys. Rev. Res.}\ }\textbf {\bibinfo {volume} {2}},\ \bibinfo {pages} {013057} (\bibinfo {year} {2020})}\BibitemShut {NoStop}%
\bibitem [{\citenamefont {Pontin}\ \emph {et~al.}(2020)\citenamefont {Pontin}, \citenamefont {Bullier}, \citenamefont {Toro\ifmmode~\check{s}\else \v{s}\fi{}},\ and\ \citenamefont {Barker}}]{PhysRevResearch.2.023349}%
  \BibitemOpen
  \bibfield  {author} {\bibinfo {author} {\bibfnamefont {A.}~\bibnamefont {Pontin}}, \bibinfo {author} {\bibfnamefont {N.~P.}\ \bibnamefont {Bullier}}, \bibinfo {author} {\bibfnamefont {M.}~\bibnamefont {Toro\ifmmode~\check{s}\else \v{s}\fi{}}},\ and\ \bibinfo {author} {\bibfnamefont {P.~F.}\ \bibnamefont {Barker}},\ }\href@noop {} {\bibfield  {journal} {\bibinfo  {journal} {Phys. Rev. Res.}\ }\textbf {\bibinfo {volume} {2}},\ \bibinfo {pages} {023349} (\bibinfo {year} {2020})}\BibitemShut {NoStop}%
\bibitem [{\citenamefont {Donadi}\ \emph {et~al.}(2021{\natexlab{a}})\citenamefont {Donadi}, \citenamefont {Piscicchia}, \citenamefont {Del~Grande}, \citenamefont {Curceanu}, \citenamefont {Laubenstein},\ and\ \citenamefont {Bassi}}]{donadi2021novel}%
  \BibitemOpen
  \bibfield  {author} {\bibinfo {author} {\bibfnamefont {S.}~\bibnamefont {Donadi}}, \bibinfo {author} {\bibfnamefont {K.}~\bibnamefont {Piscicchia}}, \bibinfo {author} {\bibfnamefont {R.}~\bibnamefont {Del~Grande}}, \bibinfo {author} {\bibfnamefont {C.}~\bibnamefont {Curceanu}}, \bibinfo {author} {\bibfnamefont {M.}~\bibnamefont {Laubenstein}},\ and\ \bibinfo {author} {\bibfnamefont {A.}~\bibnamefont {Bassi}},\ }\href@noop {} {\bibfield  {journal} {\bibinfo  {journal} {The European Physical Journal C}\ }\textbf {\bibinfo {volume} {81}},\ \bibinfo {pages} {1} (\bibinfo {year} {2021}{\natexlab{a}})}\BibitemShut {NoStop}%
\bibitem [{\citenamefont {Donadi}\ \emph {et~al.}(2021{\natexlab{b}})\citenamefont {Donadi}, \citenamefont {Piscicchia}, \citenamefont {Curceanu}, \citenamefont {Di{\'o}si}, \citenamefont {Laubenstein},\ and\ \citenamefont {Bassi}}]{donadi2021underground}%
  \BibitemOpen
  \bibfield  {author} {\bibinfo {author} {\bibfnamefont {S.}~\bibnamefont {Donadi}}, \bibinfo {author} {\bibfnamefont {K.}~\bibnamefont {Piscicchia}}, \bibinfo {author} {\bibfnamefont {C.}~\bibnamefont {Curceanu}}, \bibinfo {author} {\bibfnamefont {L.}~\bibnamefont {Di{\'o}si}}, \bibinfo {author} {\bibfnamefont {M.}~\bibnamefont {Laubenstein}},\ and\ \bibinfo {author} {\bibfnamefont {A.}~\bibnamefont {Bassi}},\ }\href@noop {} {\bibfield  {journal} {\bibinfo  {journal} {Nature Physics}\ }\textbf {\bibinfo {volume} {17}},\ \bibinfo {pages} {74} (\bibinfo {year} {2021}{\natexlab{b}})}\BibitemShut {NoStop}%
\bibitem [{\citenamefont {Arnquist}\ \emph {et~al.}(2022)\citenamefont {Arnquist} \emph {et~al.}}]{PhysRevLett.129.080401}%
  \BibitemOpen
  \bibfield  {author} {\bibinfo {author} {\bibfnamefont {I.~J.}\ \bibnamefont {Arnquist}} \emph {et~al.} (\bibinfo {collaboration} {Majorana Collaboration}),\ }\href@noop {} {\bibfield  {journal} {\bibinfo  {journal} {Phys. Rev. Lett.}\ }\textbf {\bibinfo {volume} {129}},\ \bibinfo {pages} {080401} (\bibinfo {year} {2022})}\BibitemShut {NoStop}%
\bibitem [{\citenamefont {Diosi}(1987)}]{diosi1987universal}%
  \BibitemOpen
  \bibfield  {author} {\bibinfo {author} {\bibfnamefont {L.}~\bibnamefont {Diosi}},\ }\href@noop {} {\bibfield  {journal} {\bibinfo  {journal} {Physics Letters A}\ }\textbf {\bibinfo {volume} {120}},\ \bibinfo {pages} {377} (\bibinfo {year} {1987})}\BibitemShut {NoStop}%
\bibitem [{\citenamefont {Penrose}(1996)}]{penrose1996gravity}%
  \BibitemOpen
  \bibfield  {author} {\bibinfo {author} {\bibfnamefont {R.}~\bibnamefont {Penrose}},\ }\href@noop {} {\bibfield  {journal} {\bibinfo  {journal} {General relativity and gravitation}\ }\textbf {\bibinfo {volume} {28}},\ \bibinfo {pages} {581} (\bibinfo {year} {1996})}\BibitemShut {NoStop}%
\bibitem [{\citenamefont {Pearle}(1989)}]{pearle1989combining}%
  \BibitemOpen
  \bibfield  {author} {\bibinfo {author} {\bibfnamefont {P.}~\bibnamefont {Pearle}},\ }\href@noop {} {\bibfield  {journal} {\bibinfo  {journal} {Physical Review A}\ }\textbf {\bibinfo {volume} {39}},\ \bibinfo {pages} {2277} (\bibinfo {year} {1989})}\BibitemShut {NoStop}%
\bibitem [{\citenamefont {Ghirardi}\ \emph {et~al.}(1990)\citenamefont {Ghirardi}, \citenamefont {Pearle},\ and\ \citenamefont {Rimini}}]{ghirardi1990markov}%
  \BibitemOpen
  \bibfield  {author} {\bibinfo {author} {\bibfnamefont {G.~C.}\ \bibnamefont {Ghirardi}}, \bibinfo {author} {\bibfnamefont {P.}~\bibnamefont {Pearle}},\ and\ \bibinfo {author} {\bibfnamefont {A.}~\bibnamefont {Rimini}},\ }\href@noop {} {\bibfield  {journal} {\bibinfo  {journal} {Physical Review A}\ }\textbf {\bibinfo {volume} {42}},\ \bibinfo {pages} {78} (\bibinfo {year} {1990})}\BibitemShut {NoStop}%
\bibitem [{\citenamefont {Smirne}\ and\ \citenamefont {Bassi}(2015)}]{smirne2015dissipative}%
  \BibitemOpen
  \bibfield  {author} {\bibinfo {author} {\bibfnamefont {A.}~\bibnamefont {Smirne}}\ and\ \bibinfo {author} {\bibfnamefont {A.}~\bibnamefont {Bassi}},\ }\href@noop {} {\bibfield  {journal} {\bibinfo  {journal} {Scientific reports}\ }\textbf {\bibinfo {volume} {5}},\ \bibinfo {pages} {1} (\bibinfo {year} {2015})}\BibitemShut {NoStop}%
\bibitem [{\citenamefont {Bahrami}\ \emph {et~al.}(2014)\citenamefont {Bahrami}, \citenamefont {Smirne},\ and\ \citenamefont {Bassi}}]{bahrami2014role}%
  \BibitemOpen
  \bibfield  {author} {\bibinfo {author} {\bibfnamefont {M.}~\bibnamefont {Bahrami}}, \bibinfo {author} {\bibfnamefont {A.}~\bibnamefont {Smirne}},\ and\ \bibinfo {author} {\bibfnamefont {A.}~\bibnamefont {Bassi}},\ }\href@noop {} {\bibfield  {journal} {\bibinfo  {journal} {Physical Review A}\ }\textbf {\bibinfo {volume} {90}},\ \bibinfo {pages} {062105} (\bibinfo {year} {2014})}\BibitemShut {NoStop}%
\bibitem [{\citenamefont {Nobakht}\ \emph {et~al.}(2018)\citenamefont {Nobakht}, \citenamefont {Carlesso}, \citenamefont {Donadi}, \citenamefont {Paternostro},\ and\ \citenamefont {Bassi}}]{nobakht2018unitary}%
  \BibitemOpen
  \bibfield  {author} {\bibinfo {author} {\bibfnamefont {J.}~\bibnamefont {Nobakht}}, \bibinfo {author} {\bibfnamefont {M.}~\bibnamefont {Carlesso}}, \bibinfo {author} {\bibfnamefont {S.}~\bibnamefont {Donadi}}, \bibinfo {author} {\bibfnamefont {M.}~\bibnamefont {Paternostro}},\ and\ \bibinfo {author} {\bibfnamefont {A.}~\bibnamefont {Bassi}},\ }\href@noop {} {\bibfield  {journal} {\bibinfo  {journal} {Physical Review A}\ }\textbf {\bibinfo {volume} {98}},\ \bibinfo {pages} {042109} (\bibinfo {year} {2018})}\BibitemShut {NoStop}%
\bibitem [{\citenamefont {Vinante}\ \emph {et~al.}(2019)\citenamefont {Vinante}, \citenamefont {Pontin}, \citenamefont {Rashid}, \citenamefont {Toro{\v{s}}}, \citenamefont {Barker},\ and\ \citenamefont {Ulbricht}}]{vinante2019testing}%
  \BibitemOpen
  \bibfield  {author} {\bibinfo {author} {\bibfnamefont {A.}~\bibnamefont {Vinante}}, \bibinfo {author} {\bibfnamefont {A.}~\bibnamefont {Pontin}}, \bibinfo {author} {\bibfnamefont {M.}~\bibnamefont {Rashid}}, \bibinfo {author} {\bibfnamefont {M.}~\bibnamefont {Toro{\v{s}}}}, \bibinfo {author} {\bibfnamefont {P.}~\bibnamefont {Barker}},\ and\ \bibinfo {author} {\bibfnamefont {H.}~\bibnamefont {Ulbricht}},\ }\href@noop {} {\bibfield  {journal} {\bibinfo  {journal} {Physical Review A}\ }\textbf {\bibinfo {volume} {100}},\ \bibinfo {pages} {012119} (\bibinfo {year} {2019})}\BibitemShut {NoStop}%
\bibitem [{\citenamefont {Di~Bartolomeo}\ \emph {et~al.}(2023)\citenamefont {Di~Bartolomeo}, \citenamefont {Carlesso}, \citenamefont {Piscicchia}, \citenamefont {Curceanu}, \citenamefont {Derakhshani},\ and\ \citenamefont {Di\'osi}}]{PhysRevA.108.012202}%
  \BibitemOpen
  \bibfield  {author} {\bibinfo {author} {\bibfnamefont {G.}~\bibnamefont {Di~Bartolomeo}}, \bibinfo {author} {\bibfnamefont {M.}~\bibnamefont {Carlesso}}, \bibinfo {author} {\bibfnamefont {K.}~\bibnamefont {Piscicchia}}, \bibinfo {author} {\bibfnamefont {C.}~\bibnamefont {Curceanu}}, \bibinfo {author} {\bibfnamefont {M.}~\bibnamefont {Derakhshani}},\ and\ \bibinfo {author} {\bibfnamefont {L.}~\bibnamefont {Di\'osi}},\ }\href@noop {} {\bibfield  {journal} {\bibinfo  {journal} {Phys. Rev. A}\ }\textbf {\bibinfo {volume} {108}},\ \bibinfo {pages} {012202} (\bibinfo {year} {2023})}\BibitemShut {NoStop}%
\bibitem [{\citenamefont {Dania}\ \emph {et~al.}(2023)\citenamefont {Dania}, \citenamefont {Bykov}, \citenamefont {Goschin}, \citenamefont {Teller},\ and\ \citenamefont {Northup}}]{dania2023ultra}%
  \BibitemOpen
  \bibfield  {author} {\bibinfo {author} {\bibfnamefont {L.}~\bibnamefont {Dania}}, \bibinfo {author} {\bibfnamefont {D.~S.}\ \bibnamefont {Bykov}}, \bibinfo {author} {\bibfnamefont {F.}~\bibnamefont {Goschin}}, \bibinfo {author} {\bibfnamefont {M.}~\bibnamefont {Teller}},\ and\ \bibinfo {author} {\bibfnamefont {T.~E.}\ \bibnamefont {Northup}},\ }\href@noop {} {\bibfield  {journal} {\bibinfo  {journal} {arXiv preprint arXiv:2304.02408}\ } (\bibinfo {year} {2023})}\BibitemShut {NoStop}%
\bibitem [{\citenamefont {Di~Bartolomeo}\ \emph {et~al.}(2021)\citenamefont {Di~Bartolomeo}, \citenamefont {Carlesso},\ and\ \citenamefont {Bassi}}]{PhysRevD.104.104027}%
  \BibitemOpen
  \bibfield  {author} {\bibinfo {author} {\bibfnamefont {G.}~\bibnamefont {Di~Bartolomeo}}, \bibinfo {author} {\bibfnamefont {M.}~\bibnamefont {Carlesso}},\ and\ \bibinfo {author} {\bibfnamefont {A.}~\bibnamefont {Bassi}},\ }\href@noop {} {\bibfield  {journal} {\bibinfo  {journal} {Phys. Rev. D}\ }\textbf {\bibinfo {volume} {104}},\ \bibinfo {pages} {104027} (\bibinfo {year} {2021})}\BibitemShut {NoStop}%
\bibitem [{\citenamefont {Mancini}\ and\ \citenamefont {Tombesi}(1994)}]{mancini1994quantum}%
  \BibitemOpen
  \bibfield  {author} {\bibinfo {author} {\bibfnamefont {S.}~\bibnamefont {Mancini}}\ and\ \bibinfo {author} {\bibfnamefont {P.}~\bibnamefont {Tombesi}},\ }\href@noop {} {\bibfield  {journal} {\bibinfo  {journal} {Physical Review A}\ }\textbf {\bibinfo {volume} {49}},\ \bibinfo {pages} {4055} (\bibinfo {year} {1994})}\BibitemShut {NoStop}%
\bibitem [{\citenamefont {Gardiner}\ and\ \citenamefont {Zoller}(2004)}]{gardiner2004quantum}%
  \BibitemOpen
  \bibfield  {author} {\bibinfo {author} {\bibfnamefont {C.}~\bibnamefont {Gardiner}}\ and\ \bibinfo {author} {\bibfnamefont {P.}~\bibnamefont {Zoller}},\ }\href@noop {} {\emph {\bibinfo {title} {Quantum Noise}}}\ (\bibinfo  {publisher} {Springer-Verlag Berlin},\ \bibinfo {year} {2004})\BibitemShut {NoStop}%
\bibitem [{\citenamefont {Paternostro}\ \emph {et~al.}(2006)\citenamefont {Paternostro}, \citenamefont {Gigan}, \citenamefont {Kim}, \citenamefont {Blaser}, \citenamefont {B{\"o}hm},\ and\ \citenamefont {Aspelmeyer}}]{paternostro2006reconstructing}%
  \BibitemOpen
  \bibfield  {author} {\bibinfo {author} {\bibfnamefont {M.}~\bibnamefont {Paternostro}}, \bibinfo {author} {\bibfnamefont {S.}~\bibnamefont {Gigan}}, \bibinfo {author} {\bibfnamefont {M.~S.}\ \bibnamefont {Kim}}, \bibinfo {author} {\bibfnamefont {F.}~\bibnamefont {Blaser}}, \bibinfo {author} {\bibfnamefont {H.}~\bibnamefont {B{\"o}hm}},\ and\ \bibinfo {author} {\bibfnamefont {M.}~\bibnamefont {Aspelmeyer}},\ }\href@noop {} {\bibfield  {journal} {\bibinfo  {journal} {New Journal of Physics}\ }\textbf {\bibinfo {volume} {8}},\ \bibinfo {pages} {107} (\bibinfo {year} {2006})}\BibitemShut {NoStop}%
\bibitem [{\citenamefont {Millen}\ \emph {et~al.}(2020)\citenamefont {Millen}, \citenamefont {Monteiro}, \citenamefont {Pettit},\ and\ \citenamefont {Vamivakas}}]{millen2020optomechanics}%
  \BibitemOpen
  \bibfield  {author} {\bibinfo {author} {\bibfnamefont {J.}~\bibnamefont {Millen}}, \bibinfo {author} {\bibfnamefont {T.~S.}\ \bibnamefont {Monteiro}}, \bibinfo {author} {\bibfnamefont {R.}~\bibnamefont {Pettit}},\ and\ \bibinfo {author} {\bibfnamefont {A.~N.}\ \bibnamefont {Vamivakas}},\ }\href@noop {} {\bibfield  {journal} {\bibinfo  {journal} {Reports on Progress in Physics}\ }\textbf {\bibinfo {volume} {83}},\ \bibinfo {pages} {026401} (\bibinfo {year} {2020})}\BibitemShut {NoStop}%
\bibitem [{\citenamefont {Gonzalez-Ballestero}\ \emph {et~al.}(2021)\citenamefont {Gonzalez-Ballestero}, \citenamefont {Aspelmeyer}, \citenamefont {Novotny}, \citenamefont {Quidant},\ and\ \citenamefont {Romero-Isart}}]{gonzalez2021levitodynamics}%
  \BibitemOpen
  \bibfield  {author} {\bibinfo {author} {\bibfnamefont {C.}~\bibnamefont {Gonzalez-Ballestero}}, \bibinfo {author} {\bibfnamefont {M.}~\bibnamefont {Aspelmeyer}}, \bibinfo {author} {\bibfnamefont {L.}~\bibnamefont {Novotny}}, \bibinfo {author} {\bibfnamefont {R.}~\bibnamefont {Quidant}},\ and\ \bibinfo {author} {\bibfnamefont {O.}~\bibnamefont {Romero-Isart}},\ }\href@noop {} {\bibfield  {journal} {\bibinfo  {journal} {Science}\ }\textbf {\bibinfo {volume} {374}},\ \bibinfo {pages} {eabg3027} (\bibinfo {year} {2021})}\BibitemShut {NoStop}%
\bibitem [{\citenamefont {Moore}\ and\ \citenamefont {Geraci}(2021)}]{moore2021searching}%
  \BibitemOpen
  \bibfield  {author} {\bibinfo {author} {\bibfnamefont {D.~C.}\ \bibnamefont {Moore}}\ and\ \bibinfo {author} {\bibfnamefont {A.~A.}\ \bibnamefont {Geraci}},\ }\href@noop {} {\bibfield  {journal} {\bibinfo  {journal} {Quantum Science and Technology}\ }\textbf {\bibinfo {volume} {6}},\ \bibinfo {pages} {014008} (\bibinfo {year} {2021})}\BibitemShut {NoStop}%
\bibitem [{\citenamefont {Vinante}\ \emph {et~al.}(2020{\natexlab{b}})\citenamefont {Vinante}, \citenamefont {Gasbarri}, \citenamefont {Timberlake}, \citenamefont {Toro{\v{s}}},\ and\ \citenamefont {Ulbricht}}]{vinante2020testing}%
  \BibitemOpen
  \bibfield  {author} {\bibinfo {author} {\bibfnamefont {A.}~\bibnamefont {Vinante}}, \bibinfo {author} {\bibfnamefont {G.}~\bibnamefont {Gasbarri}}, \bibinfo {author} {\bibfnamefont {C.}~\bibnamefont {Timberlake}}, \bibinfo {author} {\bibfnamefont {M.}~\bibnamefont {Toro{\v{s}}}},\ and\ \bibinfo {author} {\bibfnamefont {H.}~\bibnamefont {Ulbricht}},\ }\href@noop {} {\bibfield  {journal} {\bibinfo  {journal} {Physical Review Research}\ }\textbf {\bibinfo {volume} {2}},\ \bibinfo {pages} {043229} (\bibinfo {year} {2020}{\natexlab{b}})}\BibitemShut {NoStop}%
\bibitem [{\citenamefont {Toro{\v{s}}}\ \emph {et~al.}(2017)\citenamefont {Toro{\v{s}}}, \citenamefont {Gasbarri},\ and\ \citenamefont {Bassi}}]{torovs2017colored}%
  \BibitemOpen
  \bibfield  {author} {\bibinfo {author} {\bibfnamefont {M.}~\bibnamefont {Toro{\v{s}}}}, \bibinfo {author} {\bibfnamefont {G.}~\bibnamefont {Gasbarri}},\ and\ \bibinfo {author} {\bibfnamefont {A.}~\bibnamefont {Bassi}},\ }\href@noop {} {\bibfield  {journal} {\bibinfo  {journal} {Physics Letters A}\ }\textbf {\bibinfo {volume} {381}},\ \bibinfo {pages} {3921} (\bibinfo {year} {2017})}\BibitemShut {NoStop}%
\bibitem [{\citenamefont {Adler}(2007)}]{adler2007lower}%
  \BibitemOpen
  \bibfield  {author} {\bibinfo {author} {\bibfnamefont {S.~L.}\ \bibnamefont {Adler}},\ }\href@noop {} {\bibfield  {journal} {\bibinfo  {journal} {Journal of Physics A: Mathematical and Theoretical}\ }\textbf {\bibinfo {volume} {40}},\ \bibinfo {pages} {2935} (\bibinfo {year} {2007})}\BibitemShut {NoStop}%
\bibitem [{\citenamefont {Ghirardi}\ \emph {et~al.}(1986)\citenamefont {Ghirardi}, \citenamefont {Rimini},\ and\ \citenamefont {Weber}}]{ghirardi1986unified}%
  \BibitemOpen
  \bibfield  {author} {\bibinfo {author} {\bibfnamefont {G.~C.}\ \bibnamefont {Ghirardi}}, \bibinfo {author} {\bibfnamefont {A.}~\bibnamefont {Rimini}},\ and\ \bibinfo {author} {\bibfnamefont {T.}~\bibnamefont {Weber}},\ }\href@noop {} {\bibfield  {journal} {\bibinfo  {journal} {Physical review D}\ }\textbf {\bibinfo {volume} {34}},\ \bibinfo {pages} {470} (\bibinfo {year} {1986})}\BibitemShut {NoStop}%
\bibitem [{\citenamefont {Vacchini}(2000)}]{vacchini2000completely}%
  \BibitemOpen
  \bibfield  {author} {\bibinfo {author} {\bibfnamefont {B.}~\bibnamefont {Vacchini}},\ }\href@noop {} {\bibfield  {journal} {\bibinfo  {journal} {Physical Review Letters}\ }\textbf {\bibinfo {volume} {84}},\ \bibinfo {pages} {1374} (\bibinfo {year} {2000})}\BibitemShut {NoStop}%
\end{thebibliography}%
\end{document}